\documentclass{appolb}
\usepackage{graphics}
\def \etc{{\sl etc.\/}}
\def \beq{\begin{equation}}
\def \eeq{\end{equation}}
\def \beqa{\begin{eqnarray}}
\def \eeqa{\end{eqnarray}}
\def \lms{\Lambda_{\overline{\scriptscriptstyle MS}}}
\def \alphas{\alpha_{\scriptscriptstyle S}}
\def \cfft{\chi_{\scriptscriptstyle FFT}}
\begin{document}
\title{Lattice QCD for RHIC
\thanks{Presented at the 42nd Cracow School of Theoretical Physics, Zakopane, Poland}}
\author{Sourendu Gupta
\address{Department of Theoretical Physics, Tata Institute of Fundamental
         Research, Homi Bhabha Road, Mumbai 400005, India.}}
\maketitle
\begin{abstract}
I briefly introduce the methods by which lattice QCD predictions for RHIC
are obtained. Next I deal with lattice determinations of strangeness
production and event-to-event fluctuations of conserved quantities.
I also present a new diagrammatic method for computing derivatives with
respect to chemical potentials, and conclude with discussions of some tests
of thermal perturbation theory which follow.
\end{abstract}

\section{Introduction}

Heavy-ion collisions at the RHIC have already given evidence for dense and
hot matter \cite{general}, and may lead to a discovery of the predicted
plasma phase of QCD if such predictions are made precise enough. In recent
years computations in lattice field theories have become precise enough
to confront phenomenological analyses of experimental results. Among
the many interesting results from RHIC \cite{rhic} I single out three
for comment--- indications of early thermalisation leading possibly
to hydrodynamic flow, rapid chemical saturation of strangeness, and
fluctuations from one event to another.  Each of these observations may
be related to quantities which are easily computed in finite temperature
lattice QCD.  Flow is strongly connected to the equation of state,
strangeness to the Wroblewski parameter, and fluctuations to various
susceptibilities. The equation of state has been adequately dealt with
elsewhere \cite{kanaya}, and I will restrict myself to the rest.

Interesting thermodynamical quantities can be constructed by taking the
derivative of the free energy with respect to intensive quantities.
Consider QCD with the intensive variables temperature, $T$, and
the quark chemical potentials, $\mu_u$, $\mu_d$, $\mu_s$ (for flavours
$u$, $d$ and $s$ of quarks). The first derivative of the free energy, $F$,
with respect to one of the chemical potentials is the quark number---
\beq
   \langle n_f\rangle = \frac{\partial}{\partial\mu_f}
                                         F(T,\mu_u,\mu_d,\mu_s)
     = \frac{\partial}{\partial\mu_f} \log Z(T,\mu_u,\mu_d,\mu_s),
\label{qn}\eeq
where $Z$ is the partition function.
The second derivatives are called quark number susceptibilities
\cite{milc}
\beq
   \chi_{fg} = \frac{\partial^2}{\partial\mu_f\partial\mu_g}
                                         F(T,\mu_u,\mu_d,\mu_s)
      = \frac{\partial\langle n_g\rangle}{\partial\mu_f} 
      = \frac{\partial\langle n_f\rangle}{\partial\mu_g}\,.
\label{qns}\eeq
In general such second derivatives measure microscopic fluctuations in
equilibrium. It has recently been discovered that these fluctuations,
$\chi_{fg}$, may be directly accessible in heavy-ion collisions
\cite{fluct}. Higher derivatives, which we deal with later, may be
called non-linear quark number susceptibilities, in analogy with similar
quantities in condensed matter physics.

\begin{figure}[htb]\begin{center}
   \scalebox{0.6}{\includegraphics{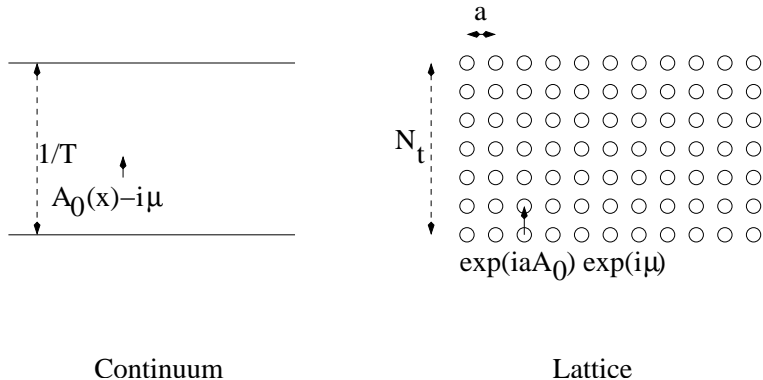}}
   \end{center}
   \caption{Finite temperature Euclidean field theory in the continuum and
      on a lattice.}
\label{fg.lattice}\end{figure}

I will describe information on these derivatives which we have obtained from
lattice simulations of QCD. These are numerical estimates of the QCD
partition function, $Z$, in the Euclidean thermal field theory context---
\beq
   Z(T,\mu_u,\mu_d,\mu_s) 
       = \int dU \prod_{f=u,d,s}\det M(m_f,\mu_f) \exp[-S(U)].
\label{zqcd}\eeq
In this formula $S(U)$ is the gauge part of the action, $M$ is the Dirac
operator for quarks of mass $m_f$ with chemical potential $\mu_f$, and
the integration is performed over all configurations of gauge fields. The
chemical potentials enter the Dirac operator as if they were constant
$U(1)$ gauge fields (see Figure \ref{fg.lattice}). In the Euclidean
formulation of finite temperature field theory the temperature enters
indirectly through the fact that the Euclidean `time' direction is of
extent $1/T$ \cite{kapusta}.

Since such integrals have the usual ultraviolet divergences of field
theory, they can be defined with a space-time lattice as a regulator.
The spacing between lattice sites in all directions is $a$.  The gauge
field $A_\mu$ associated with infinitesimal changes in position, \ie,
$\partial_\mu$, is replaced by the finite transporter $\exp(iaA_\mu)$.
The number of lattice sites in the time direction is $N_t$. At fixed
temperature this gives the relation
\beq
   a N_t = 1/T,
\label{cutoff}\eeq
This is used to eliminate the ultraviolet cutoff, $\Lambda=1/a$ from
all computations in favour of the physical scale $T$. After this
is done, the regulator must be removed by taking $\Lambda\to\infty$
while holding fixed all physical quantities. This process is called
``taking the continuum limit'', since it means that $a\to0$ at fixed
temperature by taking $N_t\to\infty$ while holding $T$ constant (see
Figure \ref{fg.lattice}).

Taking the continuum limit is exactly the same as normalizing the
field theory.  A measurement of the renormalized strong coupling at
the scale of $1/a$ flows according to the two-loop $\beta$-function of
QCD. As a result, good control over the continuum extrapolation comes
from perturbation theory, yielding \cite{alphas}
\beq
   \frac{T_c}{\lms} = \cases{1.15\pm0.05&\hskip1cm($N_f=0$),\cr
                             0.49\pm0.05&\hskip1cm($N_f=2$).}
\label{tc}\eeq
This gives us good precision in pinning down the running coupling
at any given temperature since $log(T/\lms)=\log(T/T_c)+\log(T_c/\lms)$.
$T_c/\lms$ is a reasonably easy quantity to measure because there are
nice definitions of the renormalized QCD coupling on the lattice which
show the usual logarithmic scaling without any power corrections in $a$.
Other quantities may have power corrections which need to be subtracted
before the logarithmic scaling can be seen. While this is tedious,
the great advantage of the lattice is that it allows full control over
infrared divergences which plague finite temperature field theory.

Another fact is crucial. In a lattice computation we do not determine
the integral in eq.\ (\ref{zqcd}) before taking its derivatives.  Instead,
we take the derivatives before doing the integral numerically. For example,
we notice that for any matrix $M(x)$ where each matrix element may depend
on some variable $x$
\beq
   \frac{\partial\det M(x)}{\partial x}=\det M(x) \Tr M'M^{-1},
\label{deriv}\eeq
where $M'$ denotes the matrix each term of which is the derivative of the
corresponding term of $M$. As a result,
\beq
   \langle n_i\rangle = \frac1Z\int dU \Tr M_i'M_i^{-1} \prod_f M_f
      \exp[-S(U)] = \langle\Tr M_i'M_i^{-1}\rangle.
\label{number}\eeq
The expectation value on the right is computed with a Monte Carlo
procedure which simulates the integrand of eq.\ (\ref{zqcd})
\cite{note1}. Such a procedure works only when the integrand is
non-negative. Since the Euclidean Dirac operator with chemical potential
has complex eigenvalues, the determinant is not positive definite, and
lattice Monte Carlo simulations of QCD at finite chemical potential become
tremendously hard to do. In all the work reported here we deal with the
susceptibilities evaluated at zero chemical potential. Interestingly,
they can be (and have been) used to continue lattice QCD information to
non-zero chemical potential \cite{mu}.

Finally a word about flavour symmetry breaking. If the $u$ and $d$
quark masses in nature were equal then flavour symmetry would be broken
only in electro-weak interactions. Lattice QCD computations are done in
this limit.  In reality, however, $u$ and $d$ quark masses differ. It
turns out that this breaking is almost irrelevant to thermodynamics
\cite{gavai}. The strange quark is much heavier, with a mass not much
different from $T_c$. Hence it is almost quenched close to $T_c$ but
should be treated as unquenched far above $T_c$.

\section{Fluctuations}

\subsection{Lattice Measurements}

I will introduce some notation. The usual baryon chemical potential is
$\mu_0=(\mu_u+\mu_d+\mu_s)/3$. The chemical potential associated with the
isospin quantum number is $\mu_3=(\mu_u-\mu_d)/2$. The corresponding
number densities are $\langle n_0\rangle = \langle n_u + n_d +
n_s\rangle/3$ and $\langle n_3\rangle=\langle n_u - n_d\rangle/2$. These
are zero whenever the chemical potentials vanish. The susceptibilities
obtained by taking double derivatives with respect to $\mu_0$ and $\mu_3$
are written $\chi_0$ and $\chi_3$. These can be non-zero even when the
chemical potentials vanish.

\begin{figure}[htb]\begin{center}
   \scalebox{0.6}{\includegraphics{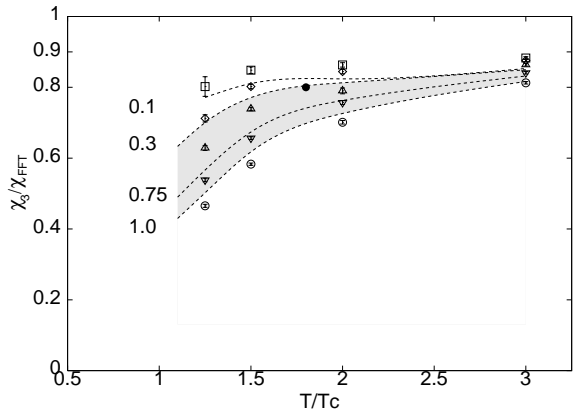}}
   \end{center}
   \caption{The susceptibility $\chi_3$ as a function of $T$ for several
       different valence quark masses, $m_v$ and lattice spacing $a=1/4T$.
       The lines summarize data for quenched QCD while the symbols are
       for data in $N_f=2$ QCD for the $m_v/T_c$ values shown. Error bars
       are mostly smaller than the symbols. The shaded region covers a
       range of quark masses appropriate to strange quarks.}
\label{fg.sus4}\end{figure}

We recently improved upon previous measurements \cite{milc,others}
of these quantities in several ways. First, by changing the size of
the spatial box within which the lattice computation is done, we have
found a range of sizes such that the box has no effect on the physical
measurement. All our subsequent measurements are in this range of box
sizes. Secondly, unlike previous computations, we have held the quark
mass, $m$, fixed in terms of physical mass scales such as GeV or $T_c$ as
we change the temperature. Previous studies had, for convenience, fixed
$ma=m/TN_t$, which meant that their quark mass changed as they scanned
across temperature. Finally, by improving the estimators of the traces,
it turns out that modern computers can allow us to reduce the error
bars in some of the susceptibilities by over 3 orders of magnitude--- an
effective gain of a factor of a over million in the statistics available
ten years ago when the last computations were performed.

For $a=1/4T$, \ie, $N_t=4$, our simulations with the quenched theory
($N_f=0$) \cite{quenched} and with two light flavours of sea quarks
($N_f=2$) \cite{dynam} showed that the susceptibility is quite different
from that for the ideal gas, $\cfft$, on the same lattice. There is a
small effect from unquenching the light sea quarks--- about 5\%. Since
this effect is so small, it seems that the effect of unquenching the
strange quark should be smaller than the statistical errors in our
measurement. By making the valence quark heavier we can therefore
investigate the dynamics of strange quarks. We will return to this
important point later.

\begin{figure}[htb]\begin{center}
   \scalebox{0.6}{\includegraphics{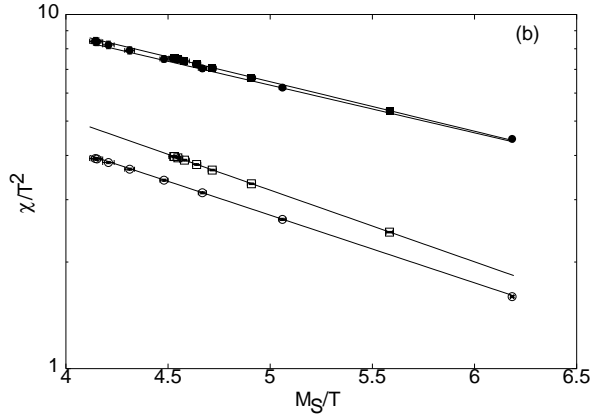}}
   \end{center}
   \caption{The dependence of $\chi_3$ and the pseudo-scalar susceptibility,
       $\chi_\pi$, on the common screening mass of the scalar and pseudo-scalar
       at $T=1.5T_c$ and $2T_c$. Error bars are smaller than the symbols.
       $\chi_\pi$ is the zero-momentum pseudo-scalar correlator \cite{suscep}.}
\label{fg.massdep}\end{figure}

The quark mass dependence of $\chi_3$ is quite nontrivial and is
interesting in itself. It has been known for a long time that $\chi_3$
can be written as the zero-momentum limit of a certain component of a
vector correlation function at finite temperature \cite{kapusta}. Now,
the breaking of Lorentz symmetry at finite temperature due to the
selection of a preferred frame (that of the heat bath) means that
angular momenta do not necessarily label the states of the system
\cite{saumen}. One of the most well-known consequences of this is the
difference between electric and magnetic polarizations of gauge bosons
at finite temperature \cite{weldon}. In any case, this broken symmetry
causes $\chi_3$ to mix with scalar/pseudo-scalar representations
\cite{irreps,dynam}. Previous work on the lattice has seen clearly that
correlations in this channel at high temperatures deviate strongly from
that of an ideal quark gas \cite{screen}. As the quark mass is changed,
this correlation function, the corresponding screening mass, and $\chi_3$
all change in response (see Figure \ref{fg.massdep}).

The flavour off-diagonal susceptibility $\chi_{ud}$ turns out to
be surprisingly small. Our measurements reveal that above $T_c$ the
dimensionless number $\chi_{ud}/T^2$ is zero to within a few parts
in $10^5$. This is a major surprise, because a recent computation in
resummed finite temperature QCD shows that this quantity should be of
the order of $\alphas^3(2\pi T)\log\alphas(2\pi T)$ and predicts that it
should be of order $10^{-3}$ \cite{bir}. A non-log contribution of order
$\alpha^3(2\pi T)$ remains to be computed, but even if this cancels the
computed term at some $T$, the range of temperatures over which results
are available is large enough that a substantial non-zero value would
still be seen. This disagreement between the lattice and perturbative
computations stand as a puzzle.

\begin{figure}[htb]\begin{center}
   \scalebox{0.6}{\includegraphics{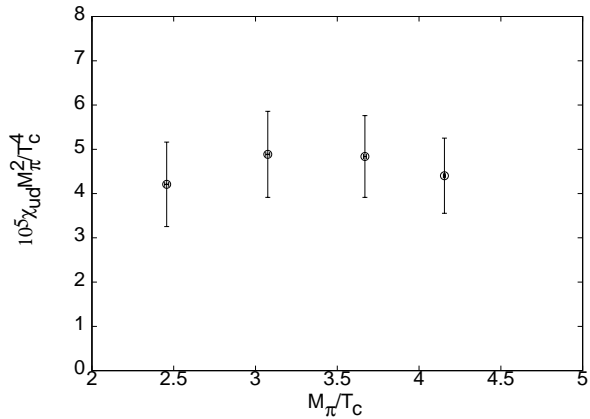}}
   \end{center}
   \caption{The dependence of $\chi_{ud}$ on $m_\pi$ at $T=0.75T_c$ in
       quenched QCD.}
\label{fg.chiud}\end{figure}

Below $T_c$ this off-diagonal susceptibility has only been investigated
in the quenched theory. It is small but clearly non-zero (see Figure
\ref{fg.chiud}).  With changing quark mass it is seen to vary roughly as
$1/m_\pi^2$ where $m_\pi$ is the pion screening mass, showing that such
fluctuations are essentially carried by pions. The connection between
the vector-vector correlator $\chi_{ud}$ and the pion below $T_c$ comes
from the fact that the contribution
\beq
   \scalebox{0.6}{\includegraphics{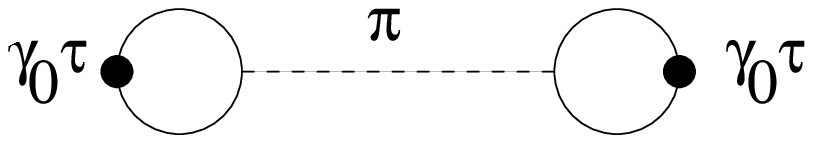}}
\label{mix}\eeq
to $\chi_{ud}$ is allowed at finite temperature, and therefore, dominates
$\rho$ exchange contributions purely kinematically. This is another
manifestation of the same physics that led to the correlation shown in
Figure \ref{fg.massdep}.  While $\chi_{ud}$ is non vanishing below $T_c$,
$\chi_3$ is consistent with zero.

Continuum extrapolation of these results have been attempted
\cite{wrob,heller,unwritten}.  It turns out that $\chi_3$ measured with
staggered Fermions have large power corrections in $a$.  As a result,
the ratio $\chi_3/\cfft$ at $T=2T_c$ decreases from $a=1/4T$ to $1/6T$
but then turns over and approaches the limit from below. The same limit
is obtained by extrapolating $\chi_3/T^2$ using the usual staggered
Fermions or an improved version, although both these extrapolations
approach the limit more smoothly. The results at $2T_c$ are in marginal
disagreement with the resummed perturbative computations of \cite{bir}
(\ie, disagrees at the 1-$\sigma$ level but agrees at 3-$\sigma$), but
become compatible with it at $T=3T_c$. $\chi_{ud}/T^2$ remains compatible
with zero at the level of a few parts in $10^{-5}$ in the continuum limit.

\subsection{Applications to phenomenology}

Two important qualitative observations emerge from the lattice computation.
First, that above $T_c$ one has non-vanishing $\chi_3$ but $\chi_{ud}$ is
zero. Second, that below $T_c$ $\chi_3$ vanishes and $\chi_{ud}$ is non-zero.
Since there are only two independent types of susceptibilities (as I show later),
all fluctuations of interest are governed by these two and their changes with
valence quark mass. 

Fluctuations of electric charge, for example, are controlled by the susceptibility
\beq
   \chi_q = \frac19\left(10\chi_3+\chi_s+\chi_{ud}-2\chi_{us}\right),
\label{chiq}\eeq
whereas baryon number fluctuations are related to
\beq
   \chi_0 = \frac19\left(4\chi_3+\chi_s+4\chi_{ud}+4\chi_{us}\right).
\label{chi0}\eeq
The following quantitative conclusions can be obtained---
\begin{enumerate}
\item For $T\gg T_c$, since $\chi_3\approx\chi_s\gg\chi_{ud}\approx\chi_{us}$,
   we have $\chi_q\approx(11/9)\chi_3$ and $\chi_0\approx(5/9)\chi_3$ so that
   the ratio $\chi_q/\chi_0\approx2$.
\item When $T>T_c$ but very close to $T_c$, since $\chi_3\gg\chi_s\gg\chi_{ud}
   \approx\chi_{us}$, we find $\chi_q\approx(10/9)\chi_3$ and $\chi_0\approx
   (4/9)\chi_3$ so that the ratio $\chi_q/\chi_0\approx2.5$.
\item For $T<T_c$ since $\chi_{ud}\propto1/m_\pi^2$ and assuming that
   $\chi_{us}\propto1/m_K^2$, since $\chi_3\approx\chi_s\approx0$, we expect
   that $\chi_q/\chi_0\approx0.25+{\cal O}(m_\pi^2/m_K^2)$.
\end{enumerate}

Under the assumptions given above, there are diametrically opposite predictions
above and below $T_c$---
\beqa
\nonumber
   \chi_0<\chi_q<\chi_s &&\qquad (T>T_c) \\
   \chi_0>\chi_q>\chi_s &&\qquad (T<T_c)
\label{pred}\eeqa
The ordering of fluctuations in baryon number \cite{gavin}, charge \cite{fluct}
and total strangeness \cite{redlich} are therefore radically different
above and below the phase transition.

\section{Strangeness production}

Strangeness abundances in heavy-ion collisions at the CERN SPS collider and the
RHIC have been analyzed extensively. There is some consensus that the observed
chemical composition is that in equilibrium close to $T_c$, and that it cannot
arise due to hadronic rescatterings \cite{strange}. One of the central quantities
that has been extracted from data is the Wroblewski parameter
\beq
   \lambda_s =
     \frac{2\langle\bar ss\rangle}{\langle\bar uu\rangle+\langle\bar dd\rangle}.
\label{wrob}\eeq
The averages on the right are defined to be the number of primary created
quark pairs of each flavour. It turns out that lattice determinations
of static equilibrium quantities can be used to predict this dynamical
quantity under some well-defined, and testable, assumptions.

As a preparatory example, consider the electrons in a metal interacting
with external fields. In a static magnetic field, $H$, at a fixed
temperature, the response is an induced magnetization whose rate of change
with the field strength is the magnetic susceptibility, $\chi(0)$.  This
is a measure of the fluctuations of spins in thermal equilibrium. On the
other hand, when an electromagnetic wave of frequency $\omega$ propagates
through the medium, it is attenuated due to dissipative phenomena
which generate many excitations in the medium.  One can describe the
dissipation through a complex susceptibility $\chi(\omega)$, describing
the response of the material to a magnetic field, $H(\omega)$ of frequency
$\omega$ \cite{note2}. Causality relates the real and imaginary parts
of $\chi(\omega)$ through a Kramers-Kr\"onig dispersion relation. From
the fluctuation-dissipation theorem it is possible to deduce that the
complex susceptibilities are proportional to the static susceptibility
if the characteristic time scales of the system are very different from
the energy scales dominating the production process \cite{dynamic}.

\begin{figure}[htb]\begin{center}
   \scalebox{0.6}{\includegraphics{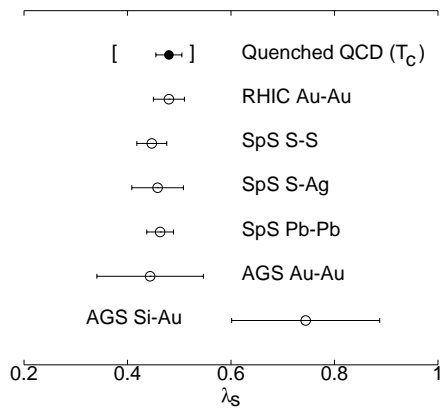}}
   \end{center}
   \caption{The Wroblewski parameter on the lattice compared with
       extraction from data \cite{cleymans}. The error bars are statistical errors.
       For the lattice extraction the bracketed interval an estimate
       of possible errors due to extrapolation to $T_c$.}
\label{fg.wrob}\end{figure}

This carries over to strangeness production. The rate of production of quark
pairs is proportional to a complex susceptibility, and hence to the static
susceptibility that we measure. This gives
\beq
   \lambda_s = \frac{2\chi_s}{\chi_u+\chi_d} = \frac{\chi_s}{\chi_u},
\label{wrobchi}\eeq
where the susceptibilities are evaluated at the temperature and chemical
potential characteristic of the collision.  In Figure \ref{fg.wrob}
we display our prediction of $\lambda_s$, from the lattice computations
already outlined, and a comparison with values extracted from experiments.
Further details, including a complete list of all assumptions and ways to
test them, can be found in \cite{wrob}.

\section{Testing perturbation theory}

We have already seen evidence of non-perturbative effects in the susceptibilities.
We extend this observation by investigating non-linear
susceptibilities. We define these as higher derivatives of the free energy---
\beq
   \chi_{fgh\cdots} = \frac{\partial^n\log Z}{\partial\mu_f\partial\mu_g\partial\mu_h\cdots}\;,
\label{nonlin}\eeq
where the flavour indices $f$, $g$, $h$, {\sl etc\/}, need not be distinct.
There is a nice and systematic way of evaluating the derivatives of $Z$. It begins
by noting that the chemical potentials $\mu_f$ appear in the partition function
of eq.\ (\ref{zqcd}) only through the quark determinant. Then, we can evaluate the
derivatives at by a chain rule starting with
\beq
   \frac{\partial\det M}{\partial\mu_f}=\det M {\cal O}_f^{(1)},\quad
   \frac{\partial^2\det M}{\partial\mu_f\partial\mu_g}=
      \det M\left[ {\cal O}_f^{(1)}{\cal O}_g^{(1)}+\delta_{fg} {\cal O}_f^{(2)}\right],
\label{derivs}\eeq
{\sl etc\/}. The operators ${\cal O}_f^{(n)}$ are defined recursively through the
relations
\beq
   {\cal O}_f^{(n)}=\frac{\partial{\cal O}_f^{(n-1)}}{\partial\mu_f},
\label{dervio}\eeq
and the concrete computational rules are---
\beq
   {\cal O}_f^{(1)}=\Tr M_f'M_f^{-1},\qquad
   \frac{\partial M_f^{-1}}{\partial\mu_f} = -M_f^{-1}M_f'M_f^{-1}.
\label{rule}\eeq
This is the complete set of rules for writing down the operator expressions for
the non-linear susceptibilities \cite{note3}.

\begin{figure}[htb]\begin{center}
   \scalebox{0.4}{\includegraphics{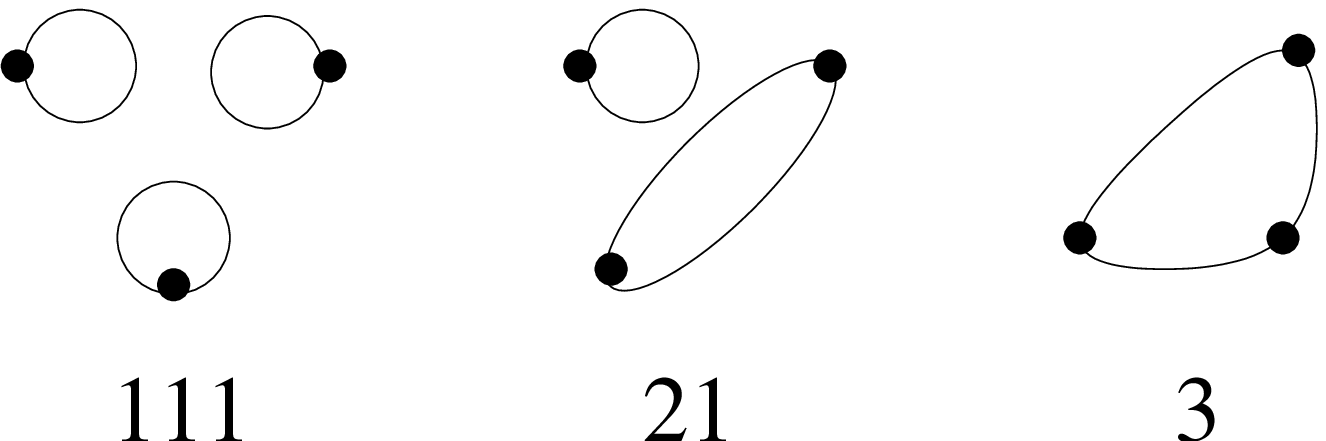}}\vskip5mm
   \scalebox{0.4}{\includegraphics{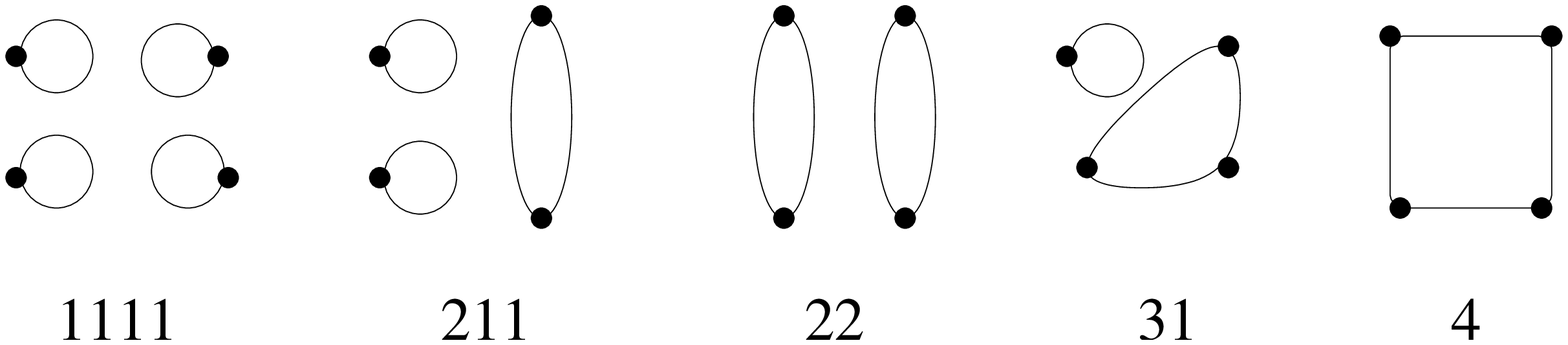}}
   \end{center}
   \caption{The operator topologies which contribute to the third and fourth
       order susceptibilities.}
\label{fg.topo}\end{figure}

In the continuum theory, since the Dirac operator contains $\mu$ linearly,
second and higher derivatives, $M_f''$ \etc, vanish.  Every $M_f'$
corresponds to an insertion of $\gamma_0\lambda_f$ (where $\lambda_f$
is a flavour generator) \cite{note4}, and each $M_f^{-1}$ is a quark
propagator. Thus the chain rule (eqs.\ \ref{derivs}--\ref{rule}) can
be written diagrammatically. The rules for a susceptibility of order
$n$ are---
\begin{enumerate}
\item Put down $n$ blobs (each corresponding to an $M_f'$, \ie, a derivative
   with respect to $\mu_f$) and label each with its flavour index.
\item Join the blobs by lines (each representing an $M_f^{-1}$) into sets of
   closed loops such that each loop contains only blobs of a single flavour.
   Count the number of ways in which each topology arises and sum them all up.
\item For degenerate flavours for $\mu_f=0$, the operator depends only on the
   topology and the flavour label on ${\cal O}_f^{(n)}$ is irrelevant. So delete
   all the flavour indices after the counting is done.
\item The operators can then be labeled only by the topology, which is specified
   completely by the number of blobs per loop and the number of such loops. Thus,
   each distinct topology is a partition of $n$.
\end{enumerate}

In Figure \ref{fg.topo} are shown the topologies that contribute to the
3rd and 4th order susceptibilities.  There are clearly two types of
operators--- one quark line connected operator for each $n$, and the
remaining quark line disconnected. Flavour off-diagonal operators are
necessarily quark-line disconnected. Since the free energy is even in
each $\mu_f$, the odd-order susceptibilities vanish for $\mu=0$, just as
do the number densities.

The rules show that $V\chi_{uu}/T=\langle{\cal O}_2+{\cal O}_{11}\rangle$
and $V\chi_{ud}/T=\langle{\cal O}_{11}\rangle$. Due to flavour symmetry,
$\chi_{uu}=\chi_{dd}$.  The number of independent physical quantities,
\ie, susceptibilities, is equal to the number of operators. Hence,
the operator expectation values are themselves physical.  At third
order $V\chi_{uud}/T=\langle {\cal O}_{111}+{\cal O}_{12}\rangle$
and $V\chi_{uuu}/T=\langle{\cal O}_{111}+ 3{\cal O}_{12}+{\cal O}_3
\rangle$. Flavour symmetry gives two different physical quantities---
$\chi_{uuu}=\chi_{ddd}$ and $\chi_{uud}=\chi_{udd}$ whereas there
are three different operators.  At fourth order, there are three
different physical quantities, which are $\chi_{uuuu}=\chi_{dddd}$,
$\chi_{uuud}=\chi_{uddd}$ and $\chi_{uudd}$, but five different
operators.  Due to flavour symmetry, the number of different $n$-th
order susceptibilities is equal to $1+n/2$ for even $n$ and $(1+n)/2$
for odd $n$, whereas the number of distinct matrix elements is the number
of partitions of $n$. For $n>2$ there are more matrix elements than
susceptibilities, and the former cannot all be physical.  The particular
fourth order susceptibility---
\beq
   \chi_{uudd}=\frac TV\langle{\cal O}_{1111} +2{\cal O}_{211}
     +{\cal O}_{22}\rangle-\chi_{uu}^2-2\chi_{ud}^2
\label{binder}\eeq
is a cumulant related to the Binder variable \cite{binder} and hence
interesting to study.

For each $n$ only one of the susceptibilities, that with only
a single flavour, contains a quark-line connected diagram. All
other susceptibilities are necessarily quark-line disconnected. We
have investigated some of these quark-line disconnected quantities
numerically.  In dynamical QCD with $N_f=2$ at temperatures
$T\ge1.5T_c$, it turns out that $\chi_{ud}/T^2$
vanishes to one part in $10^5$, and both $\chi_{uud}/T^3$
and $\chi_{uudd}/T^4$ vanish to better than one part in $10^3$.

While the quark-line disconnected diagrams are expected to vanish in an
ideal gas, in QCD they may be connected by gluon lines, and dressed by
all possible gluon and quark loops.  In \cite{bir} certain power counting
rules were developed which may be applied to operators such as these:
the main ingredient being that every loop with $n$ blobs connects to
$n_g$ electric gluon lines, where $n_g>1$ and $n_g+n$ is even.  As a
result, $\langle{\cal O}_{11}\rangle\propto g^6$ (actually $g^6\ln g$
as shown in \cite{bir} after a detailed computation).  All contributions
to the third order susceptibility vanish.  Of the diagrams contributing
to $\chi_{uudd}$, $\langle{\cal O}_{22}\rangle\propto g^4$ and gives
the leading perturbative contribution. At temperatures of $2T_c$,
for $N_f=2$, we get $\langle{\cal O}_{11}\rangle/T^2\approx0.1$,
and $\chi_{uudd}/T^4\approx0.5$. These rough perturbative estimates
can easily be modified by an order of magnitude due to subleading
logarithms and numerical coefficients. Nevertheless, the lattice results
are significantly below the perturbative estimates, and temperature
independent over a range of temperatures where the perturbative estimates
vary by a factor of 5.

This finite temperature analogue of Zweig's rule holds in a region
of temperatures away from $T_c$.  Closer to $T_c$ there there is some
evidence for non-zero values of $\chi_{ud}$ \cite{milc,others,heller}
as well as $\chi_{uudd}$.  Since these quantities measure departures
from ideal gas behaviour, they would be very interesting quantities to
study in the vicinity of the critical point of QCD.

It is a pleasure to thank the organizers for a wonderful school. I would
also like to thank Jean-Paul Blaizot and my collaborators, Rajiv Gavai
and Pushan Majumdar, for discussions.


\begin{thebibliography}{10}
\bibitem{note1}
   Notice that two conceptually different quark masses appear in
   eq. (\ref{number}). The valence quark mass, $m_v$, appears in the
   operator whose expectation value is measured--- the trace in this
   case. The sea quark mass, $m$, appears in the measure, \ie, the
   determinants.
\bibitem{note2}
   Kapusta's computation of $\chi_q$ as the self-energy of a photon propagating in
   the plasma \protect\cite{kapusta}, is precisely the static ($\omega\to0$) limit
   computation.
\bibitem{note3}
   Similar expansions and diagrammatic rules can also be written for other quantities,
   such as the chiral condensate, $\langle\bar\psi\psi\rangle=\Tr M^{-1}$, and hadron
   propagators.
\bibitem{note4}
   The non-linear terms in $\mu_f$ on the lattice are needed only for divergence
   cancellation (see \protect\cite{rajiv}), and have no effect on the physics in
   the subsequent discussion.
\end{thebibliography}
\end{document}